# Understanding Container-based Services under Software Aging: Dependability and Performance Views

Jing Bai, Xiaolin Chang, Fumio Machida, Kishor S. Trivedi

*Abstract*—**Container technology, as the key enabler behind microservice architectures, is widely applied in Cloud and Edge Computing. A long and continuous running of operating system (OS) hosting container-based services can encounter software aging that leads to performance deterioration and even causes system failures. OS rejuvenation techniques can mitigate the impact of software aging but the rejuvenation trigger interval needs to be carefully determined to reduce the downtime cost due to rejuvenation. This paper proposes a comprehensive semi-Markov-based approach to quantitatively evaluate the effect of OS rejuvenation on the dependability and the performance of a container-based service. In contrast to the existing studies, we neither restrict the distributions of time intervals of events to be exponential nor assume that backup resources are always available. Through the numerical study, we show the optimal container-migration trigger intervals that can maximize the dependability or minimize the performance of a container-based service.**

***Keywords—Container, Dependability, Semi-Markov process, Software aging, Performance***

## I. INTRODUCTION

Recent years have seen the rapid development of Container as a Service (CaaS) and Function as a Service (FaaS) utilizing the container-based virtualization in Cloud Computing [1][2], both of which have been offered by production cloud service providers, such as Amazon Web Services, IBM and Google [3]. International Data Corporation (IDC) forecasts that 80% of workload will be run in containers by 2023 [4]. Moreover, containers possess the advantages of lightweight runtime environment, very fast startup time and small memory footprint, which makes them ideal for deploying edge services in them compared with virtual machines (VMs) [5]-[7]. By 2025, 79.4 zettabytes (ZB) data is expected to be generated from 41.6 billion connected Internet of Things (IoT) devices and Edge Computing is used to process huge amounts of delay-sensitive data near data sources [8].

Application Service (AS) typically is executed in a container running on an operating system (OS) [9]. Long and continuous running of OS, which is the key component in a container-based system, can lead to software aging-induced errors [10] and then container-based service (namely, the container and AS running in it) failures. Software aging can cause the decrease in container-based service (CS) dependability and performance, further degrading user QoS (Quality of Service)/QoE (Quality of Experience) [23].

Rejuvenation techniques [32] can mitigate the impact of software aging. But the software rejuvenation procedures can cause the system downtime [11], which will affect the CS dependability and completion time (the total time required to complete CS).

Production cloud/edge service providers need to comprehensively investigate the effect of rejuvenation techniques on the CS dependability (availability and reliability) and performance (completion time) to supervise the service level agreement (SLA) fulfillment, avoiding penalization due to SLA violations [12], and then maximize benefits of them and users.

Analytical modeling is an effective approach for evaluation. Researchers have developed various analytical models for analyzing the CS dependability or performance in a virtualized system. However, some studies [14]-[21] assumed that the time intervals of all events were exponentially distributed, while the others [22]-[26] focused on evaluating one or both of CS availability, reliability and completion time. Furthermore, all these studies assumed that backup resources can not suffer from software aging. However, whether backup resources are available or not is an important factor affecting the capabilities of rejuvenation techniques. Some rejuvenation techniques, such as live container migration, require the support of backup resources. These backup resources can suffer from software aging and failure, resulting in a failed container migration and then reducing CS dependability and performance.

Additionally, it is known that the degradation of CS dependability can increase in CS completion time [13]. But the maximized CS dependability does not always mean the achievement of the optimal CS performance due to the different impact of system downtime on the dependability and completion time. Therefore, the time to trigger rejuvenation techniques, especially the time to trigger container migration after software aging (container-migration trigger interval shown in Fig. 1) must be carefully determined taking into account the trade-off between CS dependability and performance.

This paper considers a scenario where CS is executed on the Primary Host OS and can be migrated onto Backup Host OS by live container migration. For such a container-based system, we aim to quantitatively evaluate the effectiveness of rejuvenation techniques from the perspective of CS dependability and performance by applying analytical modeling approaches. To the best of our knowledge, it is the first time to quantitatively analyze the CS dependability and performance in the aforementioned system.

In summary, the main contributions of this paper are highlighted as follows:

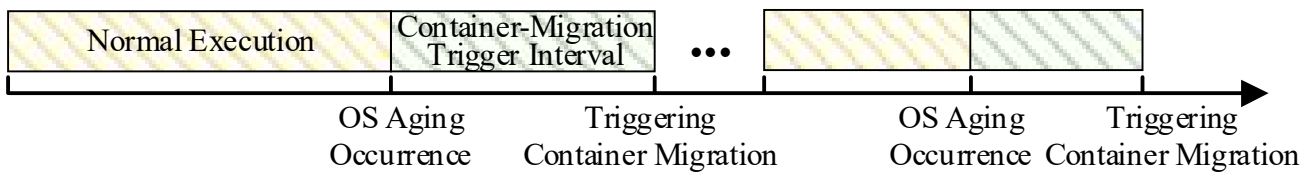

Fig. 1. Container-migration trigger interval

- We construct a semi-Markov process (SMP) model for describing the behaviors of a container-based system leveraging live container migration and OS reboot techniques for counteracting software aging. In particular, our model also depicts the behaviors of backup resources subjected to software aging and failure. That is, before live container migration technique is triggered, it should check the state of Backup Host OS.
- We derive the calculation formulas of CS completion

time, availability and reliability for studying the effectiveness of rejuvenation techniques. It is noticed that the calculation formulas of CS completion time consider the variation in CS execution rate, whose value can be different when the system is at different states. The details of system states are given in Section III.*B*.

- We conduct numerical experiments for analyzing the impact of each time interval of events following non-exponential distributions and exponential distribution on each metric and determining the optimal container-migration trigger interval. Experimental results show that each metric is sensitive to the type of failure time distributions. Our experiment results also reveal that the corresponding container-migration trigger interval is not the same when each metric is optimal.

The rest of the paper is organized as follows. Section II reviews related work. We present the SMP model for analyzing CS dependability and performance in Section III. Section IV presents the results of numerical experiments. Section V gives the conclusion and discusses the future work.

## II. Related Work

This section reviews the existing studies on model-based, event transition-based and measurement-based approaches to evaluate the dependability and/or completion time. Note that simulation and analytical model are two solution techniques in the solution phase of the model-based approaches, and they can cross-verify each other to make the evaluation results convincing.

Okamura *et al.* [14] investigated the availability of a virtualized system under two different rejuvenation policies. Torquato *et al.* [15] constructed the Stochastic Reward Nets (SRN) models and Reliability Block Diagram (RBD) to analyze the availability of different virtualized data center architectures. They further in [16] evaluated the reliability and availability of a system deploying live VM migration under varying workload based on the continuous time Markov chain (CTMC) model. Tola *et al.* [17] studied the effect of network factors on the availability of NFV (network function virtualization)-enabled service based on the minimal-cut sets and Stochastic Activity Networks (SANs). They further in [18] constructed a SAN model to capture the behaviors of different container-based NFV-MANO (management and orchestration) deployment configurations and analyzed the NFV-MANO availability. Yang *et al.* [19] proposed a CTMC model to study the reliability and availability of a repairable system. Sebastio *et al.* [20] conducted the availability analysis under different container deployment configurations based on SRN and Fault Tree (FT). The authors in [21] analyzed the availability of two container-based softwarized IP multimedia subsystem deployments, homogeneous and co-located.

Machida *et al.* [22] analyzed the effectiveness of software life-extension from the perspective of job completion time and service availability based on the SMP model. However, they did not consider the degraded performance at the aging state. The authors in [23] used the SMP model to capture the aging behaviors of virtual machine monitor (VMM) and rejuvenation behaviors by live VM migration. They evaluated the job completion time and service availability. They in [24] analyzed the optimal inspection time interval for maximizing CS availability based on Markov regenerative process (MRGP). They further evaluated the resilience, reliability and security of vehicle platooning service in [25] and analyzed the imapact of service function aging on the service dependability in [26].

There are three major differences between our work and [14]-[26]:

- *Difference 1. The systems modeled in [14]-[26] are different from our system.* The studies of [14]-[21] assumed that the time intervals of all events followed exponential distributions. The studies of [14]-[26] assumed that backup resources did not encounter software aging. In contrast to the existing studies, our work relaxes this assumption and allows the time intervals of all events to follow any type of distribution. In addition, backup resources can encounter software aging and failure in the container-based system considered in this paper.
- *Difference 2. The model construction of this paper is more comprehensive.* In the system considered in this paper, not only the added component behaviors are to be modeled, but also the additional influence among the components in the target system must be modeled. These lead to the increasing number of system states and transitions between system states.
- *Difference 3. The system is analyzed from multiple perspectives.* The studies [14]-[26] analyzed merely one or two aspects of the availability, reliability and completion time. However, we evaluate the effectiveness of rejuvenation techniques from the perspective of CS completion time, avaliability and reliability in this paper. In particular, when the completion time is calculated, more failures are discussed, which makes the results more accurate.

0 summarizes the comparison of the existing works about model-based approaches in virtualized systems. 'System' column denotes the virtualization type considered in the corresponding paper. 'Modeling Backup Resources Behaviors' column denotes whether backup resources can encounter software aging and failure in the corresponding paper. 'Model' column denotes the modelling technique employed. 'Solution Type' column indicates the type of model solution. 'Transient' indicates the transient analysis used to study the short-term behaviors [27]. 'Steady-state' indicates the steady-state analysis used to study long-term behaviors [27]. 'Analyzing The Impact of Event Time Distribution on Each Metric' column denotes whether to analyze the impact of each time interval of events following different distributions on each metric in the corresponding paper. 'Evaluation Metric' column denotes the metrics evaluated in the corresponding paper. 'Simulation' column denotes whether simulation is carried to solve the metrics or verify the approximate accuracy of the models and formulas.

Recently event transition-based approaches were developed to study the performance of time-dependent systems. The authors assessed the task completion probability of system under software aging and rejuvenation in [30]. They in [31] analyzed a real-time software system deploying both full and partial rejuvenations, and then investigated the optimal state-based rejuvenation policy in [32]. This approach was more suitable for transient system with real-time tasks that had to meet a certain completion deadline. Besides model-based and event transition-based approaches, researches explored measurement-based approaches. Oliveira *et al.* [28] evaluated the impact of software aging on the Docker Daemon and Driver by creating and periodically restarting a large number of containers and using the exec feature of Docker. However, these artificial malicious operations can be avoided. In addition, containers are an abstraction at the application layers and all containers running on a host share the host OS's kernel. Therefore, we focus on the effect of OS aging on CS dependability and performance in this paper. Note that our work can evaluate CS completion time, availability and reliability, which is complementary to these studies and thus helps service providers to provide better CS.

## III. SMP Model for Container-based System

First, this section introduces the container-based system considered in this paper. Second, we present the definitions of system states and assumptions. At last, the SMP model and the formulas of calculating CS availability, reliability and completion time are presented.

### A. *System Description*

Fig. 2 gives the container-based system architecture studied in this paper. It is composed of three components: Management Host, Primary Host and Backup Host. Primary Host includes an active container executed on the OS. Backup Host is used to support live container migration. Two types of rejuvenation techniques are considered. When OS aging is detected, live container migration is triggered. When OS failure is detected, OS is rebooted after fixing. Management Host monitors the behaviors of OS on each Host by Monitoring Tool and decides whether and which rejuvenation techniques are triggered by Analysis Tool.

Only the OSes of Primary Host and Backup Host can encounter software aging and failure. Fig. 3 shows the processing flow of a CS in the container-based system that counteracts software aging and failure by applying rejuvenation techniques. At the beginning, the CS is executed on Primary Host, while Backup Host is at idle state. We assume that OS aging can be detected immediately by monitoring memory usages [28]. In the subsequent operation process, the OS executing the CS can suffer from software aging and failure. The details are as follows.

- If the OS aging is detected during the CS processing, Monitoring Tool immediately examines the state of the Backup Host OS.
- If Backup Host OS is at the aging state, the OS is rebooted immediately so as to be ready to support live container migration. After rebooting Backup Host OS, live container migration is triggered to make Backup Host take over the CS execution. See the green dotted lines and dotted frames in Fig. 3.
- If Backup Host OS crashes, the OS is being fixed and then rebooted in order to support live container migration. After fixing and rebooting Backup Host OS, live container migration is triggered to make Backup Host take over the CS execution. See the purple dashed lines and dashed frames in Fig. 3.
- If Backup Host OS does not encounter software aging or failure, live container migration is triggered to make Backup Host take over the CS execution. See the orange dash dotted lines in Fig. 3.

After Backup Host starts the CS execution, Primary Host OS is rebooted. Requests and established sessions are usually not lost during live container migration [29]. Live container migration technique ensures that the CS can continue to be executed from the preempted point. Namely, the CS execution follows a preemptive-resume (PRS) discipline [22]. We assume that the OS aging time is much greater than the OS fixing time, OS reboot time and container-migration trigger interval.

If the OS executing CS crashes due to software aging, according to the dependencies of the AS, container and OS, the CS execution stops. After the completion of OS fixing, the OS, container, and CS are rebooted/restarted in sequence. See the blue dash double-dotted lines and dash double-dotted frames in Fig. 3. We assume that the holding times of all events follow general distributions.

### B. *State and Variable Definitions*

A 2-tuple index $(i_1, j_1)$ is defined to denote a system state. Here, $j_1$ and $i_1$ denote the states of Backup Host OS and Primary Host OS, respectively. There are seven states: Healthy, Idle, Aging, Migration, Failed, Reboot and Fixing, denoted by H, I, A, M, F, R and S, respectively. The meaning of each state is given as follows:

- State H (Healthy). OS is robust and CS is executed on it normally. OS rejuvenation techniques can bring OS back to this state.
- State I (Idle). CS is not executed on this OS.
- State A (Aging). OS at this state can work but OS aging will cause its performance to degrade. This state can help our model capture CS performance variation when OS aging occurs.
- State M (Migration). CS is ready to move from one host to another via live container migration.
- State F (Failed). At this state, OS is unavailable, which is caused by OS failure due to software aging.
- State R (Reboot). At this state, OS is rebooted.
- State S (Fixing). At this state, OS is fixed and rebooted.

There are 12 meaningful system states. Meaningless system states can be ignored. Taking system state (M, F) for example, the active container is not migrated from Primary Host to Backup Host when Backup Host OS crashes. Therefore, system state (M, F) is meaningless. TABLE III gives 12 meaningful system states.

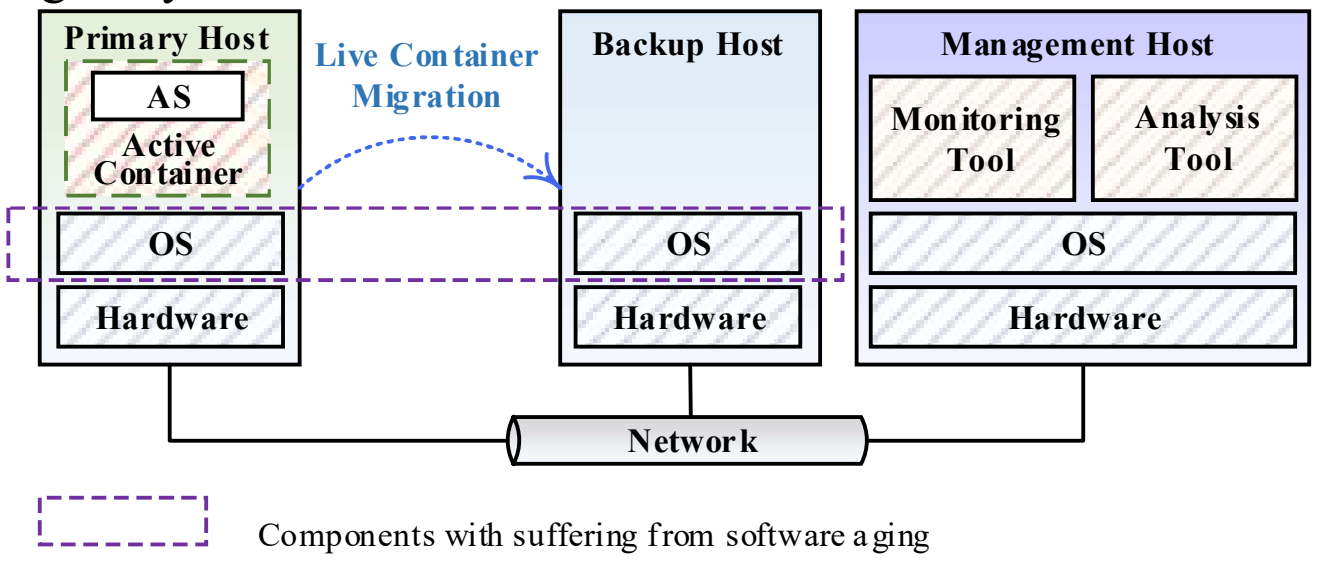

Fig. 2. System architecture

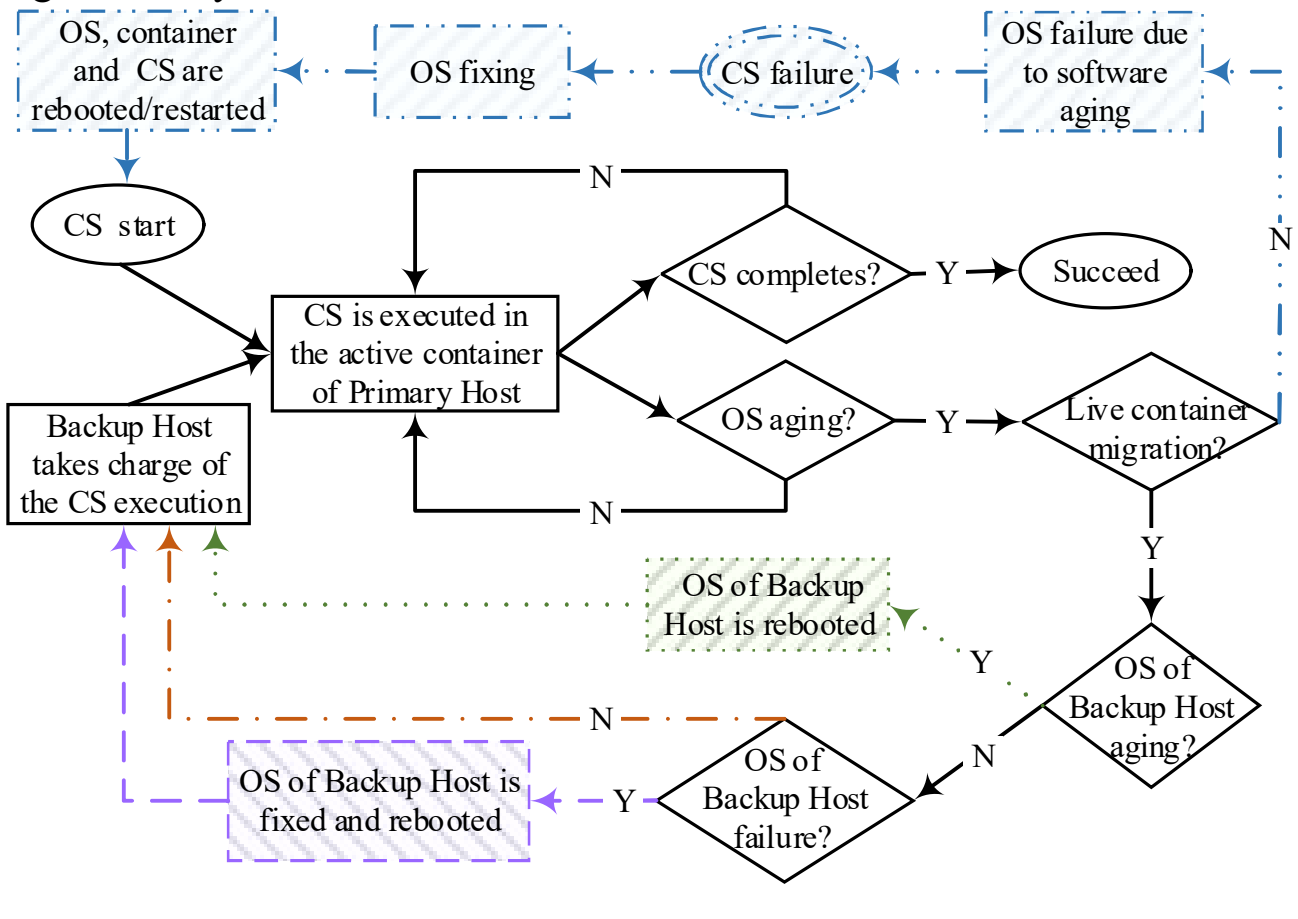

Fig. 3. The processing flow of a CS which encounters OS aging and failure

TABLE I.

TABLE II. COMPARISONS OF THE EXISTING WORKS ABOUT MODEL-BASED APPROACHES IN VIRTUALIZED SYSTEMS

| No. | System | | Modeling Backup Resources Behaviors | Model | Solution Type | Analyzing The Impact of Event Time Distribution on Each Metric | Evaluation Metric | | | Simulation |
|---|---|---|---|---|---|---|---|---|---|---|
| | VM-based | Container-based | | | | | Availability | Reliability | Completion Time | |
| [14][15] | √ | X | X | CTMC | Steady-state | X | √ | X | X | X |
| [16] | √ | X | X | CTMC | Steady-state Transient | X | √ | √ | X | X |
| [17] | √ | X | X | CTMC | Steady-state | X | √ | X | X | √ |
| [18] | X | √ | X | CTMC | Steady-state | X | √ | X | X | √ |
| [19] | √ | X | X | CTMC | Steady-state Transient | X | √ | √ | X | X |
| [20] | X | √ | X | CTMC FT | Steady-state | X | √ | X | X | √ |
| [21] | X | √ | X | CTMC | Steady-state | X | √ | X | X | X |
| [22][23] | √ | X | X | SMP | Steady-state | X | √ | X | √ | X |
| [24] | √ | X | X | MRGP | Steady-state | X | √ | X | X | X |
| [25] | X | √ | X | SMP | Steady-state Transient | X | √ | √ | X | √ |
| [26] | X | √ | X | SMP | Steady-state Transient | √ | √ | √ | X | √ |
| Ours | X | √ | √ | SMP | Steady-state | √ | √ | √ | √ | √ |

## C. SMP Model

The description of the previous sections indicates that the system behaviors at the transition epochs satisfy Markov property and the sojourn time at each system state follows non-exponential distribution. Therefore, SMP can be used for capturing behaviors that the container-based system encounters software aging and uses OS rejuvenation techniques for rejuvenation. Fig. 4 gives the SMP model. TABLE IV shows the definition of distributions used in Fig. 4. $X = \{x_0, x_1, x_2, x_3, x_4, ...\}$ is defined to denote the sequence of system states of this stochastic process at the corresponding Markov renewal moments $T = \{t_0, t_1, t_2, t_3, t_4, ...\}$. $X = \{x_0, x_1, x_2, x_3, x_4, ...\}$ is the embedded discrete time Markov chain (DTMC) [33].

## D. CS Availability Analysis

This section presents the calculation process of CS availability. We use $\mathbf{L_0}$-$\mathbf{L_{11}}$ to represent the system states (See TABLE III). The details are as follows.

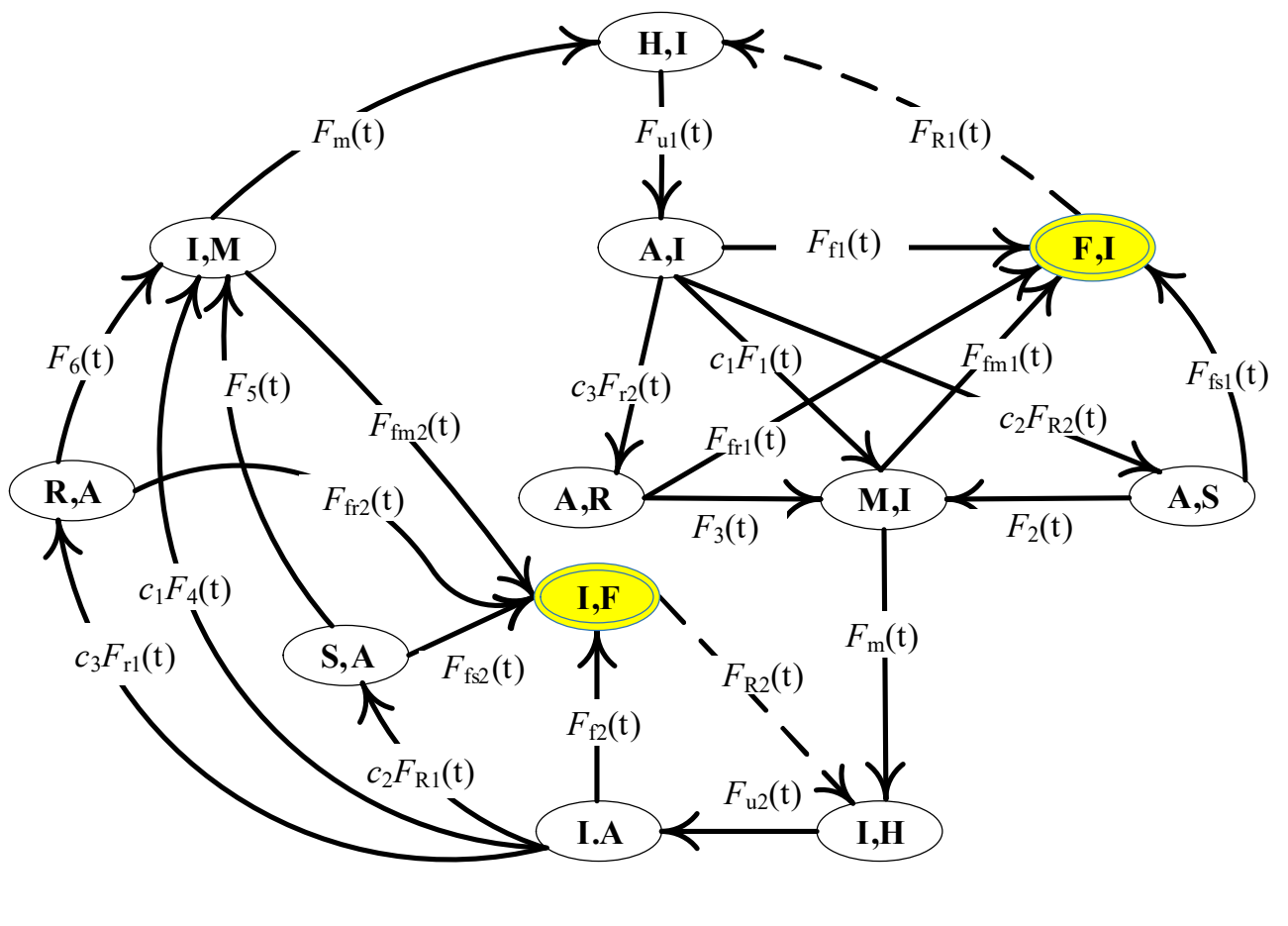

Fig. 4. SMP model

TABLE III. MEANINGFUL STATE DEFINITION

| No. | System State | State of Primary Host OS | State of Backup Host OS | Is the System Available? |
|---|---|---|---|---|
| $L_0$ | (H,I) | Healthy | Idle | Yes |
| $L_1$ | (I,A) | Idle | Aging | Yes |
| $L_2$ | (M,I) | Migration | Idle | Yes |
| $L_3$ | (A,R) | Aging | Reboot | Yes |
| $L_4$ | (S,A) | Fixing | Aging | Yes |
| $L_5$ | (R,A) | Reboot | Aging | Yes |
| $L_6$ | (A,S) | Aging | Fixing | Yes |
| $L_7$ | (I,H) | Idle | Healthy | Yes |
| $L_8$ | (A,I) | Aging | Idle | Yes |
| $L_9$ | (I,M) | Idle | Migration | Yes |
| $L_{10}$ | (F,I) | Failed | Idle | No |
| $L_{11}$ | (I,F) | Idle | Failed | No |

Firstly, the kernel matrix $\mathbf{K_{CS}}(t)$ is constructed as shown in Fig. 5 and we characterize the embedded DTMC of the SMP the embedded DTMC of the SMP by applying the one-step transition probability matrix (TPM) $\mathbf{P_{CS}} = \lim_{t\to\infty} \mathbf{K_{CS}}(t)$. The non-null elements of $\mathbf{P_{CS}}$ are given in TABLE V. Secondly, we solve the linear system of equations $\mathbf{V_{CS}}$=$\mathbf{V_{CS}}\mathbf{P_{CS}}$ subject to $\mathbf{V_{CS}} e^T = 1$ for obtaining steady-state probability vector $\mathbf{V_{CS}}$ of the embedded DTMC. Thirdly, we derive the mean sojourn time at system state $\mathbf{L_i}$, by applying $h_{\mathrm{L_i}} = \int_0^\infty (1 - H_{\mathrm{L_i}}(t))dt$, where $H_{\mathrm{L_i}}(t)$ is the sojourn time distribution at system state $\mathbf{L_i}$. The mean sojourn times are given in TABLE VI. Then, the steady-state probability $\pi_{\mathrm{L_i}}$ at system state $\mathbf{L_i}$ can be obtained by applying $\pi_{\mathrm{L_i}} = v_{\mathrm{L_i}} h_{\mathrm{L_i}} / (\sum_n v_{\mathrm{L_j}} h_{\mathrm{L_j}})$ according to [33]. Finally, the CS availability $A_{\mathbf{sa}}$ can be calculated by applying the sum of the steady-state probabilities of available states, denoted by $A_{sa} = \pi_{\mathrm{L_0}} + \pi_{\mathrm{L_1}} + \pi_{\mathrm{L_2}} + \pi_{\mathrm{L_3}} + \pi_{\mathrm{L_4}} + \pi_{\mathrm{L_5}} + \pi_{\mathrm{L_6}} + \pi_{\mathrm{L_7}} + \pi_{\mathrm{L_8}} + \pi_{\mathrm{L_9}}$.

TABLE IV. DEFINITION OF VARIABLES

| Symbol | Definition | Distribution |
|---|---|---|
| $T_{u1}$ ($T_{u2}$) | A random variable with cumulative distribution function (CDF) $F_{u1}(t)$ ($F_{u2}(t)$) denoting the holding time of Primary Host OS (Backup Host OS) from Healthy state to Aging state. (Aging time) | General distribution |
| $T_{f1}$ ($T_{f2}$) | A random variable with CDF $F_{f1}(t)$ ($F_{f2}(t)$) denoting the holding time of Primary Host OS (Backup Host OS) from Aging state to Failed state when Backup Host OS is at Idle state. (Failure time) | General distribution |
| $T_{fm1}$ ($T_{fm2}$) | A random variable with CDF $F_{fm1}(t)$ ($F_{fm2}(t)$) denoting the holding time of Primary Host OS (Backup Host OS) from Aging state to Failed state when Backup Host OS is at Migration state. (Failure time) | General distribution |
| $T_{fs1}$ ($T_{fs2}$) | A random variable with CDF $F_{fs1}(t)$ ($F_{fs2}(t)$) denoting the holding time of Primary Host OS (Backup Host OS) from Aging state to Failed state when Backup Host OS is at Fixing state. (Failure time) | General distribution |
| $T_{fr1}$ ($T_{fr2}$) | A random variable with CDF $F_{fr1}(t)$ ($F_{fr2}(t)$) denoting the holding time of Primary Host OS (Backup Host OS) from Aging state to Failed state when Backup Host OS is at Reboot state. (Failure time) | General distribution |
| $T_{r1}$ ($T_{r2}$) | A random variable with CDF $F_{r1}(t)$ ($F_{r2}(t)$) denoting the holding time of Primary Host OS (Backup Host OS) from Failed state to Healthy state. (Fixing time) | General distribution |
| $T_{R1}$ ($T_{R2}$) | A random variable with CDF $F_{R1}(t)$ ($F_{R2}(t)$) denoting the holding time of Primary Host OS (Backup Host OS) rebooting after software aging. (Reboot time) | General distribution |
| $T_M$ | A random variable with CDF $F_m(t)$ denoting the holding time of live container migration. (Container-migration time) | General distribution |
| $T_1$ | A random variable with CDF $F_1(t)=u(t\text{-a}_1)$ denoting the holding time of system state from (A,I) to (M,I). | Unit step function |
| $T_2$ | A random variable with CDF $F_2(t)=u(t\text{-a}_2)$ denoting the holding time of system state from (A,S) to (M,I). | Unit step function |
| $T_3$ | A random variable with CDF $F_3(t)=u(t\text{-a}_3)$ denoting the holding time of system state from (A,R) to (M,I). | Unit step function |
| $T_4$ | A random variable with CDF $F_4(t)=u(t\text{-a}_4)$ denoting the holding time of system state from (I,A) to (I,M). | Unit step function |
| $T_5$ | A random variable with CDF $F_5(t)=u(t\text{-a}_5)$ denoting the holding time of system state from (S,A) to (I,M). | Unit step function |
| $T_6$ | A random variable with CDF $F_6(t)=u(t\text{-a}_6)$ denoting the holding time of system state from (R,A) to (I,M). | Unit step function |
| $c_1$ | The probability that Backup Host OS is healthy when Primary Host OS is detected to suffer from software aging. | - |
| $c_2$ | The probability that aging occurs to Backup Host OS when Primary Host OS is detected to suffer from software aging. | - |
| $c_3$ | The probability that Backup Host OS fails when Primary Host OS is detected to suffer from software aging. | - |

$$\mathbf{K}_{\mathbf{CS}}(t)=\begin{pmatrix}
0&0&0&0&0&0&0&0&k_{L_0L_8}(t)&0&0&0\\
0&0&0&0&k_{L_1L_4}(t)&k_{L_1L_5}(t)&0&0&0&k_{L_1L_9}(t)&0&k_{L_1L_{11}}(t)\\
0&0&0&0&0&0&0&k_{L_2L_7}(t)&0&0&k_{\mathbf{L_2L_{10}}}(t)&0\\
0&0&k_{L_3L_2}(t)&0&0&0&0&0&0&0&k_{\mathbf{L_3L_{10}}}(t)&0\\
0&0&0&0&0&0&0&0&0&k_{L_4L_9}(t)&0&k_{L_4L_{11}}(t)\\
0&0&0&0&0&0&0&0&0&k_{L_5L_9}(t)&0&k_{L_5L_{11}}(t)\\
0&0&0&0&0&0&0&0&0&0&k_{\mathbf{L_6L_{10}}}(t)&0\\
0&k_{L_7L_1}(t)&0&0&0&0&0&0&0&0&0&0\\
0&0&k_{L_8L_2}(t)&k_{L_8L_3}(t)&0&0&k_{L_8L_6}(t)&0&0&0&k_{L_8L_{10}}(t)&0\\
k_{\mathbf{L_9L_0}}(t)&0&0&0&0&0&0&0&0&0&0&k_{\mathbf{L_9L_{11}}}(t)\\
k_{\mathbf{L_{10}L_0}}(t)&0&0&0&0&0&0&0&0&0&0&0\\
0&0&0&0&0&0&0&k_{\mathbf{L_{11}L_7}}(t)&0&0&0&0
\end{pmatrix}$$

Fig. 5. The kernel matrix $\mathbf{K}_{\mathbf{CS}}(t)$

TABLE V. THE NON-NULL ELEMENTS OF $\mathbf{P}_{\text{CS}}$

| | | |
|---|---|---|
| $p_{\mathbf{L_0L_8}}=1$ | $p_{\mathbf{L_1L_4}}=\int_0^\infty \overline{F_{f2}(x)}\,\overline{c_1F_4(x)}\,\overline{c_3F_{r1}(x)}dc_2F_{R1}(x)$ | $p_{\mathbf{L_1L_5}}=\int_0^\infty \overline{F_{f2}(x)}\,\overline{c_1F_4(x)}\,\overline{c_2F_{R1}(x)}dc_3F_{r1}(x)$ |
| $p_{\mathbf{L_1L_{11}}}=\int_0^\infty \overline{c_3F_{r1}(x)}\,\overline{c_1F_4(x)}\,\overline{c_2F_{R1}(x)}dF_{f2}(x)$ | $p_{\mathbf{L_1L_9}}=1-p_{\mathbf{L_1L_5}}-p_{\mathbf{L_1L_4}}-p_{\mathbf{L_1L_{11}}}$ | $p_{\mathbf{L_2L_7}}=\int_0^\infty \overline{F_{fm1}(x)}dF_m(x)$ |
| $p_{\mathbf{L_2L_{10}}}=\int_0^\infty \overline{F_m(x)}dF_{fm1}(x)$ | $p_{\mathbf{L_3L_{10}}}=\int_0^\infty \overline{F_3(x)}dF_{fr1}(x)$ | $p_{\mathbf{L_3L_2}}=1-p_{\mathbf{L_3L_{10}}}$ |
| $p_{\mathbf{L_4L_{11}}}=\int_0^\infty \overline{F_5(x)}dF_{fs2}(x)$ | $p_{\mathbf{L_4L_9}}=1-p_{\mathbf{L_4L_{11}}}$ | $p_{\mathbf{L_5L_{11}}}=\int_0^\infty \overline{F_6(x)}dF_{fr2}(x)$ |
| $p_{\mathbf{L_5L_9}}=1-p_{\mathbf{L_5L_{11}}}$ | $p_{\mathbf{L_6L_{10}}}=\int_0^\infty \overline{F_2(x)}dF_{fs1}(x)$ | $p_{\mathbf{L_6L_2}}=1-p_{\mathbf{L_6L_{10}}}$ |
| $p_{\mathbf{L_7L_1}}=1$ | $p_{\mathbf{L_8L_3}}=\int_0^\infty \overline{F_{f1}(x)}\,\overline{c_2F_{R2}(x)}\,\overline{c_1F_1(x)}dc_3F_{r2}(x)$ | $p_{\mathbf{L_8L_6}}=\int_0^\infty \overline{F_{f1}(x)}\,\overline{c_3F_{r2}(x)}\,\overline{c_1F_1(x)}dc_2F_{R2}(x)$ |
| $p_{\mathbf{L_8L_{10}}}=\int_0^\infty \overline{c_2F_{R1}(x)}\,\overline{c_3F_{r2}(x)}\,\overline{c_1F_1(x)}dF_{f1}(x)$ | $p_{\mathbf{L_8L_2}}=1-p_{\mathbf{L_8L_3}}-p_{\mathbf{L_8L_6}}-p_{\mathbf{L_8L_{10}}}$ | $p_{\mathbf{L_9L_0}}=\int_0^\infty \overline{F_m(t)}dF_{fm2}(t)$ |
| $p_{\mathbf{L_9L_{11}}}=1-p_{\mathbf{L_9L_0}}$ | $p_{\mathbf{L_{10}L_0}}=1$ | $p_{\mathbf{L_{11}L_7}}=1$ |

a. $\overline{cF(*)}=1-cF(*)$ and $\overline{F(*)}=1-F(*)$.

TABLE VI. THE MEAN SOJOURN TIME

| | | |
|---|---|---|
| $h_{\mathbf{L}_1} = \int_0^{\infty} \overline{F_{\mathrm{u1}}(t)} dt$ | $h_{\mathbf{L}_1} = \int_0^{\infty} \overline{F_{\mathrm{f2}}(t)}\,\overline{c_1 F_4(t)}\,\overline{c_2 F_{\mathrm{R1}}(t)}\,\overline{c_3 F_{\mathrm{r1}}(t)} dt$ | $h_{\mathbf{L}_2} = \int_0^{\infty} \overline{F_{\mathrm{fm1}}(t)}\,\overline{F_{\mathrm{m}}(t)} dt$ |
| $h_{\mathbf{L}_3} = \int_0^{\infty} \overline{F_3(t)}\,\overline{F_{\mathrm{fr1}}(t)} dt$ | $h_{\mathbf{L}_4} = \int_0^{\infty} \overline{F_5(t)}\,\overline{F_{\mathrm{fs2}}(t)} dt$ | $h_{\mathbf{L}_5} = \int_0^{\infty} \overline{F_6(t)}\,\overline{F_{\mathrm{fr2}}(t)} dt$ |
| $h_{\mathbf{L}_6} = \int_0^{\infty} \overline{F_2(t)}\,\overline{F_{\mathrm{fs1}}(t)} dt$ | $h_{\mathbf{L}_7} = \int_0^{\infty} \overline{F_{\mathrm{u2}}(t)} dt$ | $h_{\mathbf{L}_8} = \int_0^{\infty} \overline{F_{\mathrm{f1}}(t)}\,\overline{c_1 F_1(t)}\,\overline{c_2 F_{\mathrm{R2}}(t)}\,\overline{c_3 F_{\mathrm{r2}}(t)} dt$ |
| $h_{\mathbf{L}_9} = \int_0^{\infty} \overline{F_{\mathrm{fm1}}(t)}\,\overline{F_{\mathrm{m}}(t)} dt$ | $h_{\mathbf{L}_{10}} = \int_0^{\infty} \overline{F_{\mathrm{R1}}(t)} dt$ | $h_{\mathbf{L}_{11}} = \int_0^{\infty} \overline{F_{\mathrm{R2}}(t)} dt$ |

## E. CS Reliability Analysis

This section describes the process of analyzing CS reliability in terms of MTTF, which is a classical yardstick for assessing reliability [34]. MTTF can be regard as the expected time to failure for a no-fixing system. “No-fixing” means that fixing operation is not performed after OS failure. Therefore, we first re-construct SMP model introduced in Section III.*C* as SMP model with absorbing states by removing the dotted line in Fig. 4. Since the system does not recover after failure, the SMP contains absorbing states. All system states with Failed state are regarded as absorbing states (See the system states with yellow color in Fig. 4). The kernel matrix $\mathbf{K}_{\mathbf{MCS}}(t)$ is constructed for the SMP model with absorbing states. The TPM $\mathbf{P}_{\mathbf{MCS}}$ can be calculated by applying $\mathbf{P}_{\mathbf{MCS}} = \lim_{t\to\infty} \mathbf{K}_{\mathbf{MCS}}(t)$ to characterize the embedded DTMC of the SMP model with absorbing states.

The $\mathbf{P}_{\mathbf{MCS}}$ in the shape of Equation (1) can be obtained.

$$\mathbf{P}_{\mathbf{MCS}} = \begin{pmatrix} \mathbf{M} & \mathbf{c}^{\mathrm{T}} \\ \mathbf{0} & \mathbf{P}_{\mathbf{F}} \end{pmatrix} \quad (1)$$

$\mathbf{M}$ is a sub-stochastic matrix with (*n*-*m*) by (*n*-*m*) partition of matrix $\mathbf{P}_{\mathbf{MCS}}$ and describes the transitions between non-absorbing states. *m* is the number of absorbing states, each of which needs a self-loop with probability 1. $\mathbf{P}_{\mathbf{F}}$ is a sub-stochastic matrix with the *m* by *m* partition of matrix $\mathbf{P}_{\mathbf{MCS}}$ and describes the transition between absorbing states. $\mathbf{c}^{\mathrm{T}}$ is a sub-stochastic matrix with (*n*-*m*) by *m* and it is used to represent the transition probabilities from any states $\mathbf{L}_{\mathbf{j_a}}$ to the absorbing states. The expected number of visits $V_{\mathbf{L}_{\mathbf{j_a}}}$ to system state $\mathbf{L}_{\mathbf{j_a}}$ until absorption is given by:

$$V_{\mathbf{L}_{\mathbf{j_a}}} = \alpha_{\mathbf{L}_{\mathbf{j_a}}} + \sum_{\mathbf{i}=0}^{\mathbf{n}-1} V_{\mathbf{L}_{\mathbf{i}}} p_{\mathbf{L}_{\mathbf{i}}\mathbf{L}_{\mathbf{j_a}}} \quad (2)$$

where $\alpha_{\mathbf{L}_{\mathbf{j_a}}}$ is the initial probability of system state $\mathbf{L}_{\mathbf{j_a}}$ and $p_{\mathbf{L}_{\mathbf{i}}\mathbf{L}_{\mathbf{j_a}}}$ denotes the element of $\mathbf{M}$, given in TABLE V. The formula of calculating CS MTTF can be written as $\mathrm{MTTF} = V_{\mathbf{L}_0}h_{\mathbf{L}_0} + V_{\mathbf{L}_1}h_{\mathbf{L}_1} + V_{\mathbf{L}_2}h_{\mathbf{L}_2} + V_{\mathbf{L}_3}h_{\mathbf{L}_3} + V_{\mathbf{L}_4}h_{\mathbf{L}_4} + V_{\mathbf{L}_5}h_{\mathbf{L}_5} + V_{\mathbf{L}_6}h_{\mathbf{L}_6} + V_{\mathbf{L}_7}h_{\mathbf{L}_7} + V_{\mathbf{L}_8}h_{\mathbf{L}_8} + V_{\mathbf{L}_9}h_{\mathbf{L}_9}$, where $h_{\mathbf{L}_{\mathbf{j_a}}}$ is the mean sojourn time at system state $\mathbf{L}_{\mathbf{j_a}}$ and its calculation formula is the same as that in Section III.*D*.

## F. CS Performance Analysis

This section gives the calculation process of CS completion time. $C(x)$ denotes the amount of time for the completion of CS execution. We assume that the work requirement for the CS execution is *x* work units where a work unit is processed in an hour in the execution environment. If the CS suffers from a failure at time instant *h* ( $h > 0$ ), it is restarted. The calculation of CS completion time is detailed in the following.

Based on Fig. 4, we assume that CS starts its execution from (H,I). There are two cases as follows.

- The probability of a CS failure event on Primary Host is $b_1$. If $h$ is less than $x$ (the failure may occur before container migration or after container migration from Primary Host), $C(x)$ is the sum of $C(x)$, Primary Host OS fixing time and *h*. Otherwise, $C(x) = x$.
- The probability of a CS failure event on Backup Host is $b_2$. $x_1$ is defined to denote the work units completed on Primary Host. If $h$ is less than $x - x_1$ (the failure may occur before container migration or after container migration from Backup Host), $C(x)$ is the sum of $C(x)$, Backup Host OS fixing time and *h*. Otherwise, $C(x) = x$.

Equation (3) gives the formula of the mean CS completion time,

$$E(C(x)) = -\frac{\partial \Phi_C^{\sim}(s,x)}{\partial s}\Big|_{s=\mathbf{0}} \quad (3)$$

where $\Phi_C^{\sim}(s,x)$ is the Laplace-Stieltjes transform (LST) of the CS completion time. Moreover, if the impact of the mean CS execution rate $r_1$ at Aging state and the mean CS execution rate $r_2 = 1$ at Healthy state on CS completion time is considered, CS completion time is obtained by solving Equation (3). Equations (4) and (5) are the LSTs of the CS completion times in the aforementioned two cases. We can calculate $E_{P1}(C(x))$ and $E_{P2}(C(x))$ by applying Equations (4) and (5), respectively. Then, we can solve CS completion time by applying $E_P(C(x)) = \mathrm{b}_1 E_{P1}(C(x)) + \mathrm{b}_2 E_{P2}(C(x))$.

$$
\begin{aligned}
&\tilde{\Phi}_{P1}(s,x)=e^{-s(\frac{a_1}{r_1}+\frac{x-a_1/r_1}{r_1})}\tilde{F}_{\mathrm{u1}}(s)c_2F_{\mathrm{R2}}(\frac{a_1}{r_1})\\
&\int_{\frac{a_1}{r_1}+\frac{x-a_1/r_1}{r_1}}^{\infty}dF_{\mathrm{fs1}}(h-\frac{a_1}{r_1})+e^{-s(\frac{a_1}{r_1}+\frac{x-a_1/r_1}{r_1})}\tilde{F}_{\mathrm{u1}}(s)c_3F_{\mathrm{r2}}(\frac{a_1}{r_1})\\
&\int_{\frac{a_1}{r_1}+\frac{x-a_1/r_1}{r_1}}^{\infty}dF_{\mathrm{fr1}}(h-\frac{a_1}{r_1})+e^{-s(\frac{a_1}{r_1}+\frac{x-a_1/r_1}{r_1})}(1-F_{\mathrm{f1}}(\frac{a_1}{r_1})\\
&-c_2F_{\mathrm{R2}}(\frac{a_1}{r_1})-c_3F_{\mathrm{r2}}(\frac{a_1}{r_1}))\tilde{F}_{\mathrm{u1}}(s)\int_{\frac{a_1}{r_1}+\frac{x-a_1/r_1}{r_1}}^{\infty}dF_{\mathrm{fm1}}(h-\frac{a_1}{r_1})\\
&+\tilde{G}_{\mathrm{R1}}(s)\tilde{\Phi}_{P1}(s,x)\tilde{F}_{\mathrm{u1}}(s)\int_{0}^{\frac{a_1}{r_1}}e^{-sh}dF_{\mathrm{f1}}(h)+\tilde{G}_{\mathrm{R1}}(s) \qquad (4)\\
&\tilde{\Phi}_{P1}(s,x)\tilde{F}_{\mathrm{u1}}(s)c_2F_{\mathrm{R2}}(\frac{a_1}{r_1})\int_{\frac{a_1}{r_1}}^{\frac{a_1}{r_1}+\frac{x-a_1/r_1}{r_1}}e^{-sh}dF_{\mathrm{fs1}}(h-\frac{a_1}{r_1})\\
&+\tilde{\Phi}_{P1}(s,x)(1-F_{\mathrm{f1}}(\frac{a_1}{r_1})-c_2F_{\mathrm{R2}}(\frac{a_1}{r_1})-c_3F_{\mathrm{r2}}(\frac{a_1}{r_1}))\\
&\tilde{G}_{\mathrm{R1}}(s)\tilde{F}_{\mathrm{u1}}(s)\int_{\frac{a_1}{r_1}}^{\frac{a_1}{r_1}+\frac{x-a_1/r_1}{r_1}}e^{-sh}dF_{\mathrm{fm1}}(h-\frac{a_1}{r_1})+\tilde{G}_{\mathrm{R1}}(s)\\
&\tilde{\Phi}_{P1}(s,x)\tilde{F}_{\mathrm{u1}}(s)c_3F_{\mathrm{r2}}(\frac{a_1}{r_1})\int_{\frac{a_1}{r_1}}^{\frac{a_1}{r_1}+\frac{x-a_1/r_1}{r_1}}e^{-sh}dF_{\mathrm{fr1}}(h-\frac{a_1}{r_1})
\end{aligned}
$$

$$
\begin{aligned}
&\tilde{\Phi}_{P2}(s,x)=e^{-s(\frac{t_1}{r_1}+\frac{(x-x_1)-t_1/r_1}{r_1})}\tilde{F}_{\mathrm{u2}}(s)c_2F_{\mathrm{R1}}(\frac{t_1}{r_1})\\
&\int_{\frac{t_1}{r_1}+\frac{(x-x_1)-t_1/r_1}{r_1}}^{\infty}dF_{\mathrm{fs2}}(h-\frac{t_1}{r_1})+e^{-s(\frac{t_1}{r_1}+\frac{(x-x_1)-t_1/r_1}{r_1})}c_3F_{\mathrm{r1}}(\frac{t_1}{r_1})\\
&\tilde{F}_{\mathrm{u2}}(s)\int_{\frac{t_1}{r_1}+\frac{(x-x_1)-t_1/r_1}{r_1}}^{\infty}dF_{\mathrm{fr2}}(h-\frac{t_1}{r_1})+e^{-s(\frac{t_1}{r_1}+\frac{(x-x_1)-t_1/r_1}{r_1})}\\
&\tilde{F}_{\mathrm{u2}}(s)(1-F_{\mathrm{f2}}(\frac{t_1}{r_1})-c_2F_{\mathrm{R1}}(\frac{t_1}{r_1})-c_3F_{\mathrm{r1}}(\frac{t_1}{r_1}))\\
&\int_{\frac{t_1}{r_1}+\frac{(x-x_1)-t_1/r_1}{r_1}}^{\infty}dF_{\mathrm{fm2}}(h-\frac{t_1}{r_1})+\tilde{G}_{\mathrm{R2}}(s)\tilde{\Phi}_{P1}(s,x)\tilde{F}_{\mathrm{u2}}(s)\\
&\int_{0}^{\frac{t_1}{r_1}}e^{-sh}dF_{\mathrm{f2}}(h)+\tilde{G}_{\mathrm{R2}}(s)\tilde{\Phi}_{P1}(s,x)\tilde{F}_{\mathrm{u2}}(s)c_2F_{\mathrm{R1}}(t_1) \qquad (5)\\
&\int_{\frac{t_1}{r_1}}^{\frac{t_1}{r_1}+\frac{(x-x_1)-t_1/r_1}{r_1}}e^{-sh}dF_{\mathrm{fs2}}(h-\frac{t_1}{r_1})+\tilde{G}_{\mathrm{R2}}(s)\tilde{\Phi}_{P1}(s,x)\\
&\tilde{F}_{\mathrm{u2}}(s)(1-F_{\mathrm{f2}}(\frac{t_1}{r_1})-c_2F_{\mathrm{R1}}(\frac{t_1}{r_1})-c_3F_{\mathrm{r1}}(\frac{t_1}{r_1}))\\
&\int_{\frac{t_1}{r_1}}^{\frac{t_1}{r_1}+\frac{(x-x_1)-t_1/r_1}{r_1}}e^{-sh}dF_{\mathrm{fm2}}(h-\frac{t_1}{r_1})+\tilde{G}_{\mathrm{R2}}(s)\tilde{\Phi}_{P1}(s,x)\\
&\tilde{F}_{\mathrm{u2}}(s)c_3F_{\mathrm{r1}}(\frac{t_1}{r_1})\int_{\frac{t_1}{r_1}}^{\frac{t_1}{r_1}+\frac{(x-x_1)-t_1/r_1}{r_1}}e^{-sh}dF_{\mathrm{fr2}}(h-\frac{t_1}{r_1})
\end{aligned}
$$

# IV. Dependability and Performance Evaluation

In this section, we first perform the simulations to demonstrate the approximate accuracy of our model and formulas combined with numerical experiments. Second, we investigate the impact of each time interval of events following different distributions on CS availability, reliability and performance. Finally, we analyze the aforementioned metrics over various system parameters.

## A. Experiment Configuration

The time intervals of all events in Section III can follow any type of distribution. For the purposes of performing simulations and numerical experiments, the fixing times are assumed to follow Exponential distribution and Erlang distribution, denoted as $F_{\mathrm{ri}}=\mathrm{EXP}(\gamma)$ and $F_{\mathrm{ri}}(t)=\mathrm{ERL}(\gamma,v)$, respectively. The reboot times are assumed to follow Exponential distribution and Erlang distribution, denoted as $F_{\mathrm{Ri}}(t)=\mathrm{EXP}(\lambda)$ and $F_{\mathrm{Ri}}(t)=\mathrm{ERL}(\lambda,r)$, respectively. We assume that the container-migration time follows Exponential distribution and Erlang distribution, denoted as $F_{\mathrm{m}}(t)=\mathrm{EXP}(\kappa)$ and $F_{\mathrm{m}}(t)=\mathrm{ERL}(\kappa,l)$, respectively. We assume that the aging times follow Exponential distribution, Erlang distribution, and Hypoexponential distribution, denoted as $F_{\mathrm{ui}}(t)=\mathrm{EXP}(\mu)$, $F_{\mathrm{ui}}(t)=\mathrm{ERL}(\mu,c)$ and $F_{\mathrm{ui}}(t)=\mathrm{HYPO}(\mu_1,\mu_2)$, respectively. The failure times are assumed to follow Exponential distribution, Erlang distribution, and Hypoexponential distribution, denoted as $F_{\mathrm{fi}}(t)=\mathrm{EXP}(\omega)$, $F_{\mathrm{fi}}(t)=\mathrm{ERL}(\omega,g)$ and $F_{\mathrm{fi}}(t)=\mathrm{HYPO}(\omega_1,\omega_2)$, respectively.

We give the settings of the parameters (defined in TABLE IV) under different distributions in TABLE V. '-' in the 'Distribution' column denotes that the variable does not follow the corresponding distribution. The authors in [35] conducted fault injection experiments to obtained the aging time and failure time of OS, which were influenced by the strength of the work. It is difficult to estimate exactly the aging time and failure time of OS due to the difficulty in determining the representativeness of the workload. Therefore, it is reasonable to set the failure time of 40 days and the aging time of 60 days in the experiments. The fixing time and reboot time are retrieved from previous literature [20][36][37]. The migration time is set according to the measurements in literature [38], which carried out extensive experiments on the measurement of the migration time under different scenarios. The probabilities ( $c1$, $c2$ and $c3$ ) can be obtained by counting the number of abnormal events within a fixed time period. Note that these parameters are set only to demonstrate the effectiveness of the proposed approach. The mean value of each variable is unchanged when using different distributions for numerical experiments. Simulation and numerical experiments are conducted on MAPLE [39].

TABLE VII. Default Parameter Settings

| Variable | Distribution | | | Mean |
|---|---|---|---|---|
| | *Exponential* | *Erlang* | *Hypo-exponential* | |
| $T_{\mathrm{ui}}$ | 0.0006857 | 0.0013717,2 | 0.0010816, 0.0018762 | 60 days |
| $T_{\mathrm{fi}}$ | 0.0010432 | 0.0020855,2 | 0.0013674, 0.0043860 | 40 days |
| $T_{\mathrm{ri}}$ | 1 | 2, 2 | - | 1 hour [37] |
| $T_{\mathrm{Ri}}$ | 12 | 24, 2 | - | 5 minutes [37] |
| $T_{\mathrm{M}}$ | 120.5 | 720, 6 | - | 30 seconds [38] |
| $T_1$-$T_6$ | - | - | - | 0-50 hours |
| $c_1$,$c_2$,$c_3$ | - | - | - | 0-1 |

TABLE VIII. SIMULATION AND NUMERICAL RESULTS FOR CS AVAILABILITY AND MTTF

| Container-Migration Trigger Interval | CS Availability | | | CS MTTF (hour) | | |
|---|---|---|---|---|---|---|
| | *Numerical Result* | *Simulation Result* | *95% CI* | *Numerical Result* | *Simulation Result* | *95% CI* |
| 0 | 0.99985025 | 0.999800 | [0.999738026165039, 0.999861973834961] | 6674 | 6645 | [6570,6721] |
| 10 | 0.99985051 | 0.999805 | [0.999743805588150, 0.999866194411850] | 6688 | 6675 | [6599,6751] |
| 20 | 0.99985058 | 0.999810 | [0.999749595075505, 0.999870404924495] | 6691 | 6688 | [6612,6763] |
| 30 | 0.99985059 | 0.999825 | [0.999767028060878, 0.999882971939122] | 6692 | 6716 | [6640,6792] |
| 40 | 0.99985054 | 0.999815 | [0.999755395027016, 0.999874604972984] | 6689 | 6664 | [6588,6739] |
| 50 | 0.99985044 | 0.999805 | [0.999743805588150, 0.999866194411850] | 6684 | 6649 | [6574,6725] |

TABLE IX. SIMULATION AND NUMERICAL RESULTS FOR CS COMPLETION TIME

| Container-Migration Trigger Interval | CS Completion Time (hour) | | |
|---|---|---|---|
| | *Numerical Result* | *Simulation Result* | *95% CI* |
| 0 | 1721 | 1658 | [1596,1719] |
| 50 | 1702 | 1621 | [1560,1682] |
| 100 | 1696 | 1610 | [1549,1670] |
| 150 | 1701 | 1614 | [1554,1675] |
| 200 | 1721 | 1683 | [1621,1746] |

## B. Verification of Our Proposed Model and Formulas

This section presents the comparison of simulation and numerical results. We assume that the failure times are hypoexponentially distributed and the time intervals of other events are exponentially distributed in simulations. TABLE VIII and TABLE IX give the comparison of numerical and simulation results for CS completion time, CS MTTF and CS availability. '95% CI' denotes that the simulation results are calculated using 95% confidence intervals. One sees from tables that simulation results are close to the corresponding numerical results, which verifies the approximate accuracy of our formulas and model.

## C. Impact of Different Distributions on Metrics

This section studies CS availability, MTTF, and completion time under different distributions of time intervals of events. Fig. 6, Fig. 7 and Fig. 8 show the results.

'A_ERL' denotes that the aging times are Erlang distributed and the time intervals of other events are exponentially distributed. 'A_HYPO' denotes that the aging times are hypoexponentially distributed and the time intervals of other events are exponentially distributed. 'Exponential' denotes that the time intervals of all events follow exponential distributions. 'R_ERL' denotes that the reboot times are Erlang distributed and the time intervals of other events follow exponential distributions. 'Fixing_ERL' denotes that the fixing times are Erlang distributed and the time intervals of other events are exponentially distributed. 'M_ERL' denotes that the container-migration time are Erlang distributed and the time intervals of other events follow exponential distributions. 'A_HYPO-F_HYPO-ERL' denotes that the aging times and failure times follow Hypoexponential distributions and the time intervals of other events are Erlang distributed. 'F_ERL' denotes that the failure times are Erlang distributed and the time intervals of other events follow exponential distributions. 'F_HYPO' denotes that the failure times are hypoexponentially distributed and the time intervals of other events follow exponential distributions.

We can get that:

- *Finding 1. The optimal container-migration trigger intervals change significantly when we relax the assumption of exponentially distributed failure times.* Fig. 6 (a) shows the optimal container-migration trigger intervals corresponding to the maximum availability are the 0 hour under the failure times following Exponential distribution. Fig. 6 (b) shows the optimal container-migration trigger intervals corresponding to the maximum availability are between 10 and 40 hours. Similarly, one sees from Fig. 7 and Fig. 8 that the optimal container-migration trigger intervals corresponding to the optimal MTTF and completion time under the failure times following non-exponential distribution are different from those of Exponential distribution. In particular, there is a difference of several hundred hours between CS MTTF and completion time under the failure times of Exponential distribution and those of non-exponential distribution. The results under the time intervals of all events following non-exponential distributions differed from those of exponential distributions, which illustrates the superiority of our SMP model.
- *Finding 2. The optimal container-migration trigger intervals change slightly when we relax the assumption of exponentially distributed of time intervals of other events excluding the failure times (the aging, fixing, reboot and container-migration times).* Fig. 6 (a) shows that the optimal container-migration trigger intervals corresponding to the optimal availability under the time intervals of other events excluding the failure times following non-exponential distributions are close to those of exponential distribution. We can observe similar results from Fig. 7 (a) and Fig. 8 (a) that the optimal container-migration trigger intervals corresponding to the optimal MTTF and completion time under the time intervals of other events excluding the failure times following non-exponential distributions are close to those of exponential distribution. These results illustrate that the failure time distribution is a significant factor affecting CS availability, MTTF and completion time.
- *Finding 3. The optimal container-migration trigger intervals under the failure times of Erlang distribution are close to those of Hypoexponential distribution.* Fig. 6 (b) illustrates that the optimal container-migration trigger intervals corresponding to the maximum availability under the failure times of Hypoexponential distribution are close to those of Erlang distribution. We can observe similar results can be seen from Fig. 7 (b) and Fig. 8 (b) that the optimal container-migration trigger intervals corresponding to the optimal MTTF and completion time under the failure times of Hypoexponential distribution are close to those of Erlang distribution. It can be explained that the failure rates of Erlang distribution and Hypoexponential distribution increases over time.

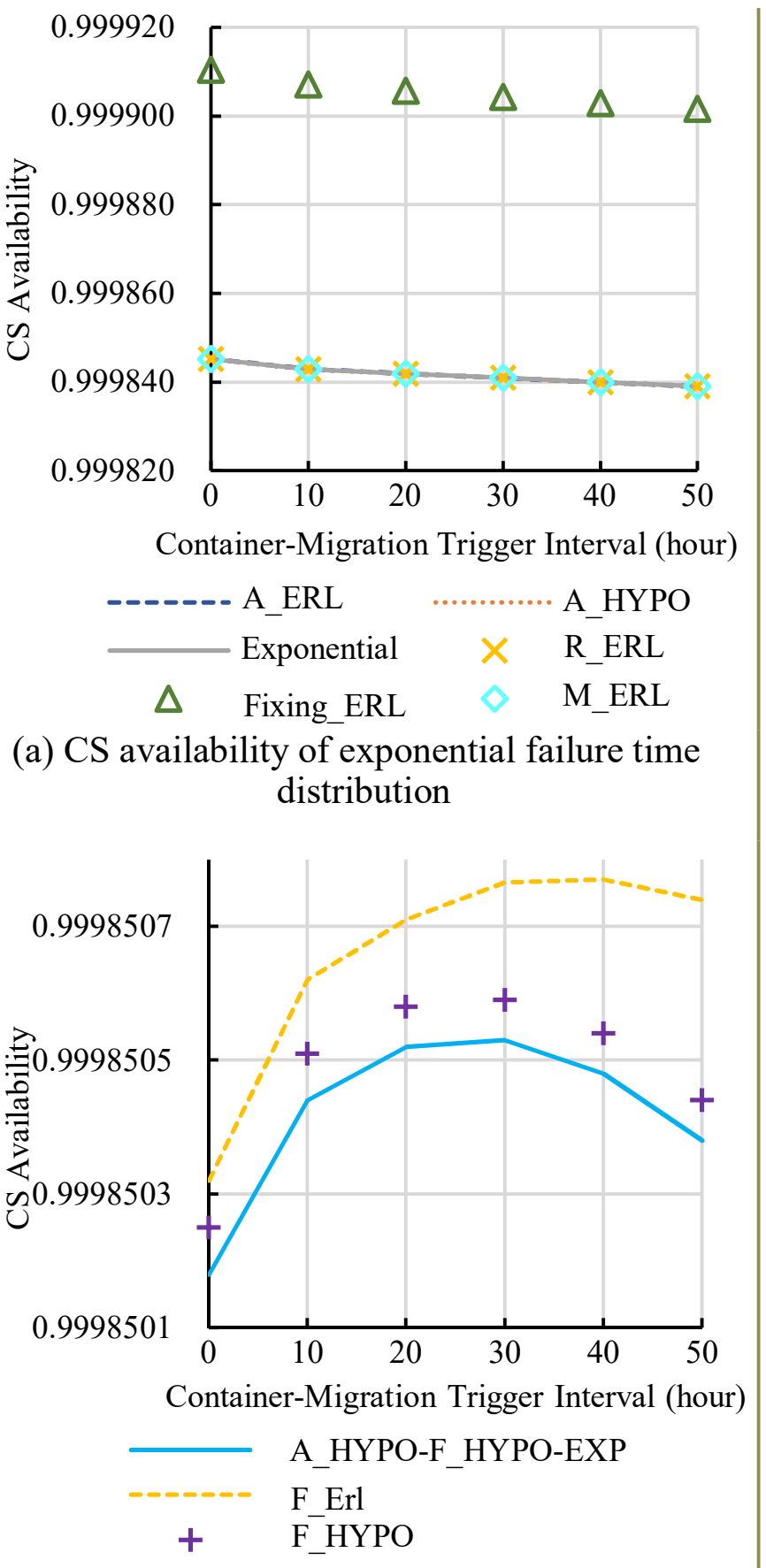

(a) CS availability of exponential failure time distribution

(b) CS availability of non-exponential failure time distribution

Fig. 6. CS availability under different distributions of time intervals of events

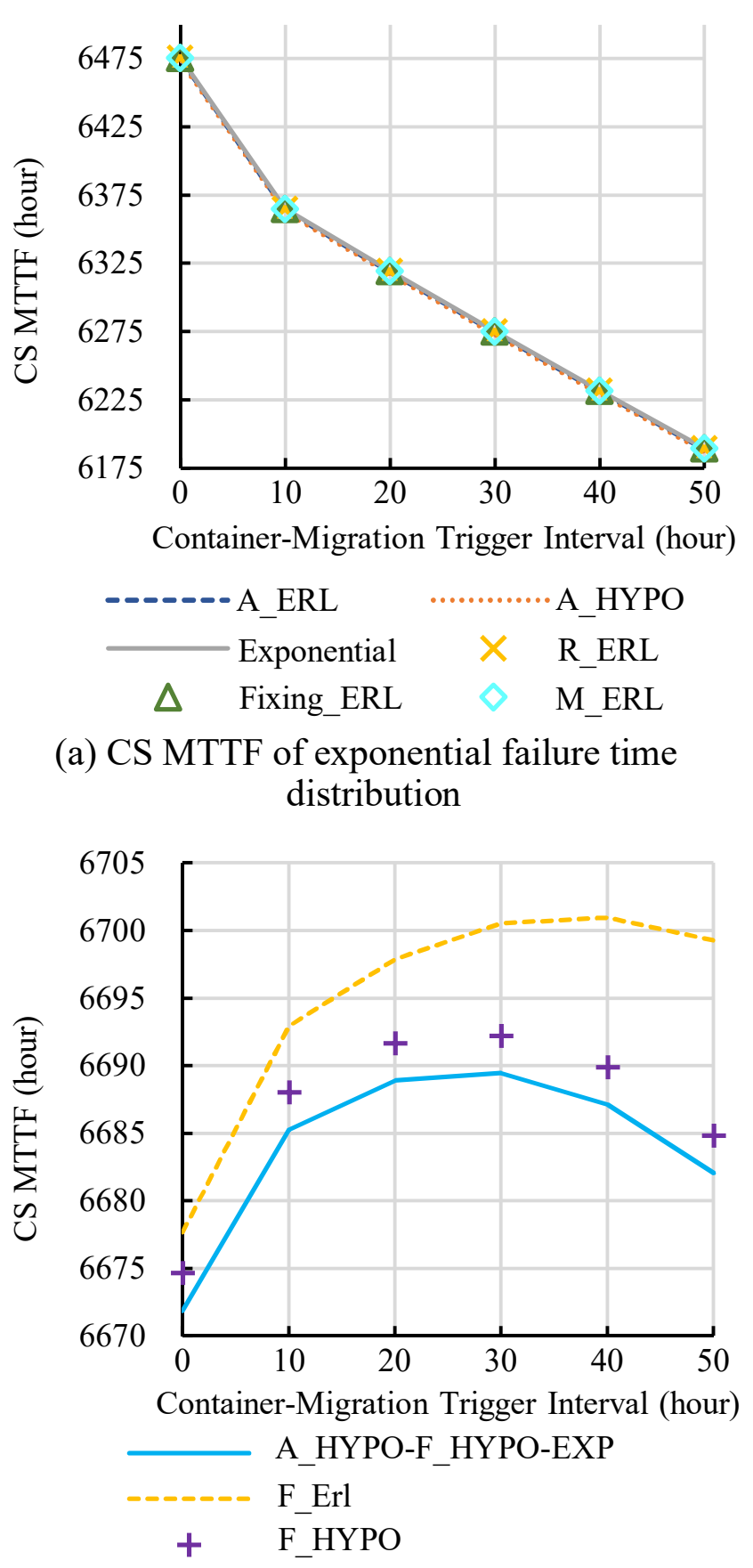

(a) CS MTTF of exponential failure time distribution

(b) CS MTTF of non-exponential failure time distribution

Fig. 7. CS MTTF under different distributions of time intervals of events

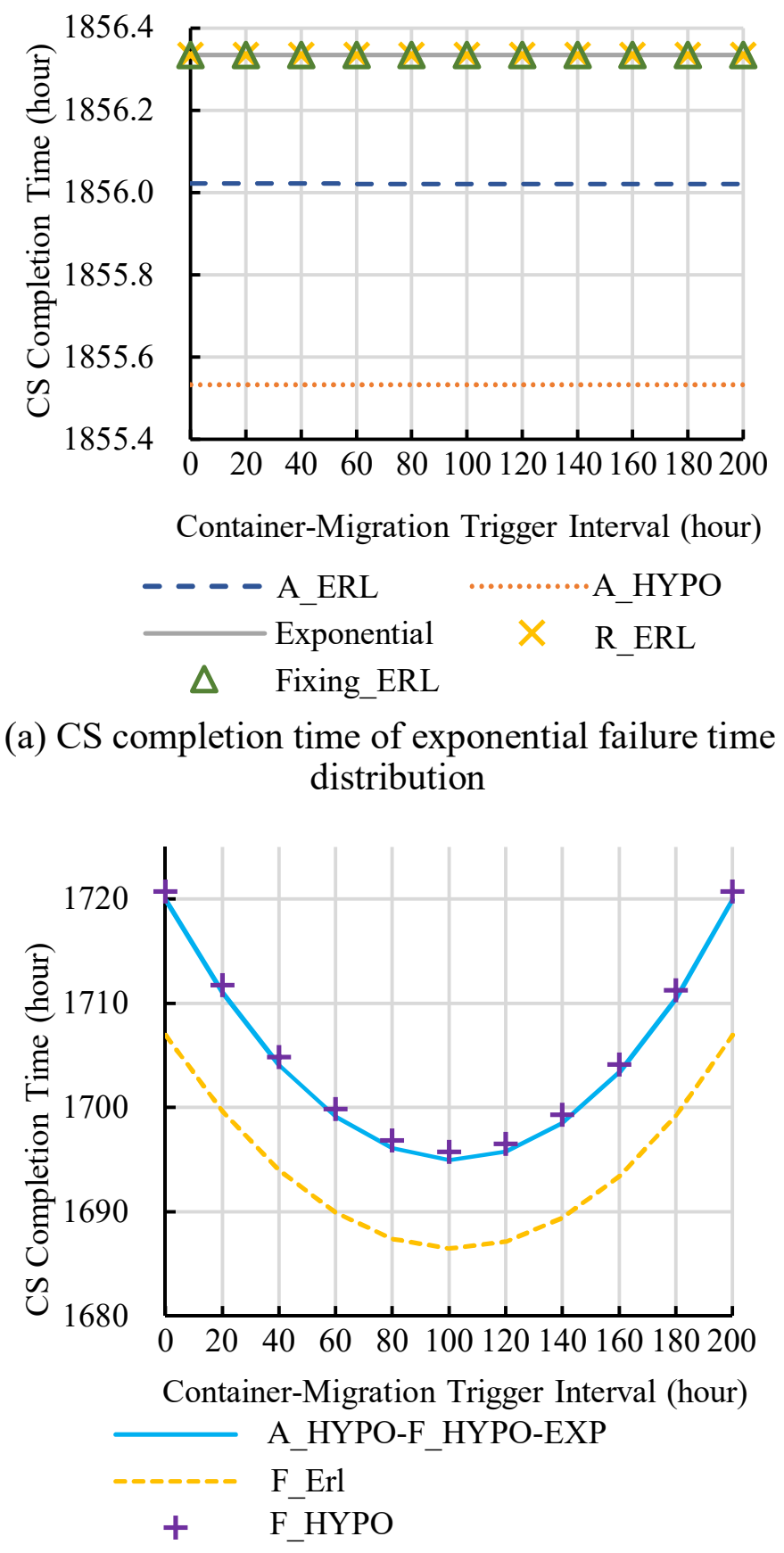

(a) CS completion time of exponential failure time distribution

(b) CS completion time of non-exponential failure time distribution

Fig. 8. CS completion time under different distributions of time intervals of events

TABLE X. OPTIMAL CS AVAILABILITY, MTTF AND COMPLETION TIME AND THEIR CORRESPONDING OPTIMAL CONTAINER-MIGRATION TRIGGER INTERVALS UNDER DIFFERENT FIXING TIMES

| Fixing Time (hour) | CS Availability | | CS MTTF | | CS Completion Time | |
|---|---|---|---|---|---|---|
| | ***Maximum CS Availability*** | ***The Corresponding Optimal Container-Migration Trigger Interval (hour)*** | ***Maximum CS MTTF (hour)*** | ***The Corresponding Optimal Container-Migration Trigger Interval (hour)*** | ***Minimum CS Completion Time (hour)*** | ***The Corresponding Optimal Container-Migration Trigger Interval (hour)*** |
| 0.8 | 0.999880417875 | 27 | 6689.4845 | 27 | 1694.9672 | 102 |
| 0.9 | 0.999864114746 | 27 | 6689.5489 | 27 | 1694.9699 | 102 |
| 1 | 0.999850529444 | 27 | 6689.6023 | 27 | 1694.9721 | 102 |
| 1.1 | 0.999833925971 | 27 | 6689.6670 | 27 | 1694.9747 | 102 |
| 1.2 | 0.999821001708 | 27 | 6689.7172 | 27 | 1694.9767 | 102 |

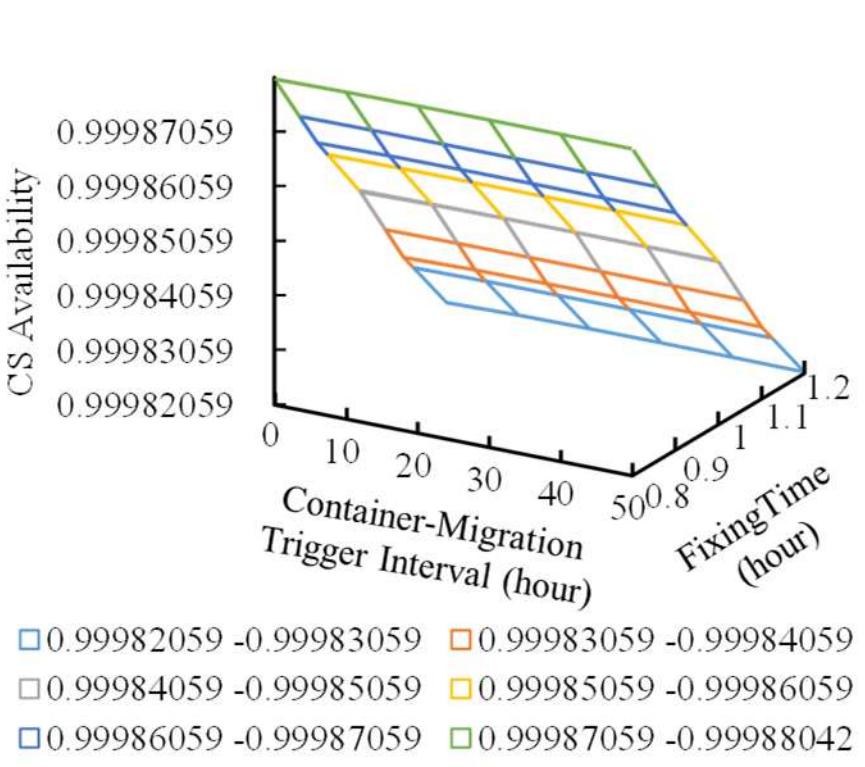

Fig. 9. CS availability over fixing time

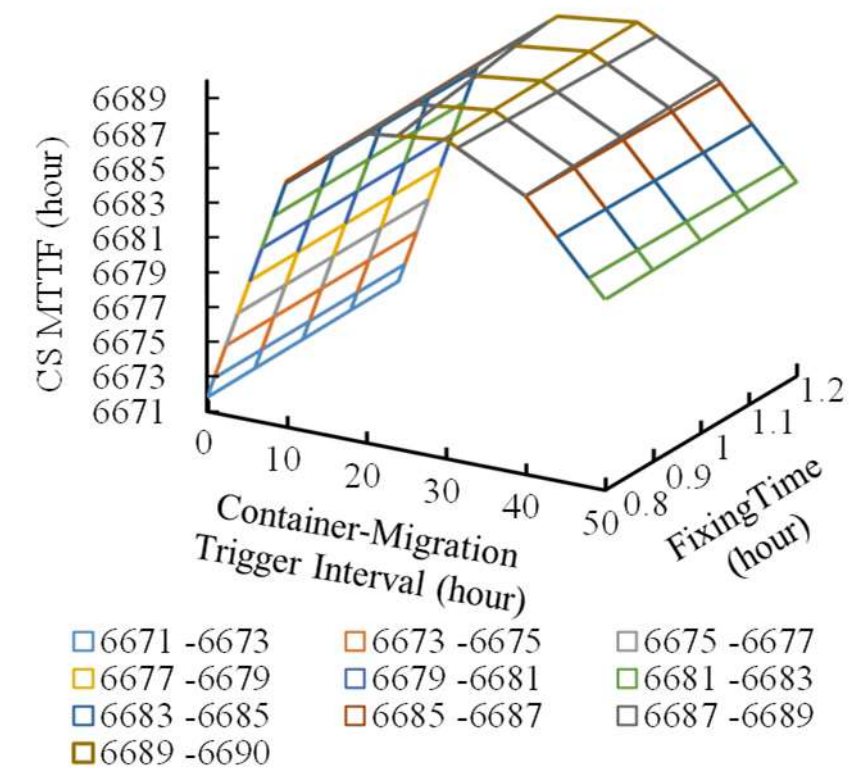

Fig. 10. CS MTTF over fixing time

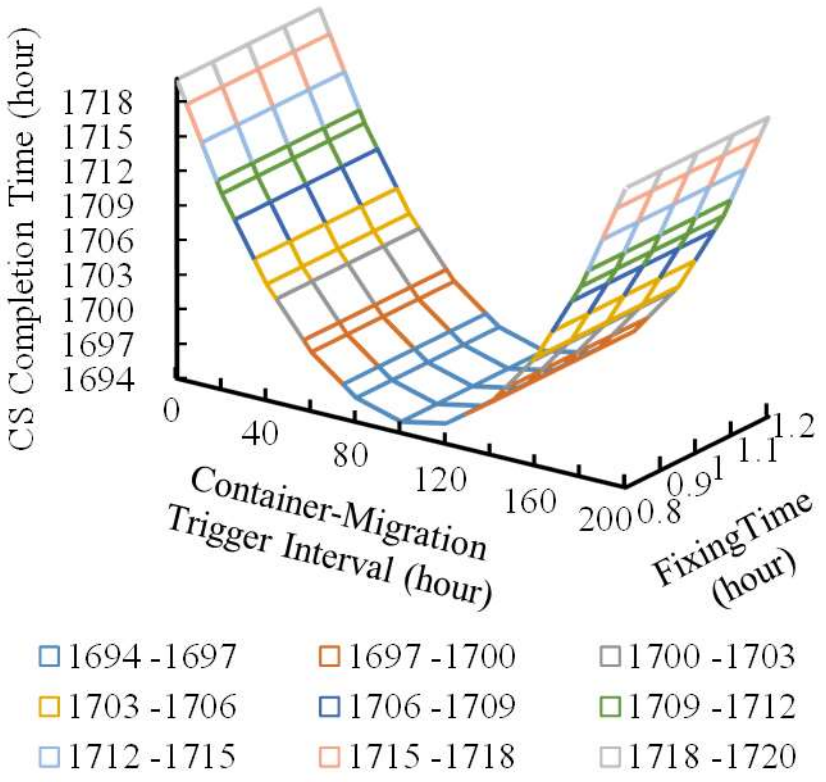

Fig. 11. CS completion time over fixing time

## D. Impact of System Parameters on Metrics

We present the relationship between container-migration trigger interval $a_1$ and metrics under different system parameters. Fig. 9, Fig. 10 and Fig. 11 show the numerical results of CS availability, MTTF and completion time under different fixing times, respectively. TABLE X gives the optimal CS availability, MTTF and completion time and their corresponding optimal container-migration trigger interval under different fixing times.

# V. Conclusions

In this paper, we proposed the SMP model to capture the behaviors of the container-based system which counteracts software aging and failure by OS rejuvenation techniques. We derive the calculation formulas of CS dependability and completion time under different container-migration trigger intervals. At last, we investigate the optimal container-migration trigger intervals for achieving the approximate minimum CS completion time and the approximate maximum CS dependability through numerical experiments, which can help production cloud/edge service providers decide when to trigger rejuvenation techniques for making a trade-off between CS performance and dependability.

# References

[1] IBM Cloud Education: FaaS (Function-as-a-Service). July 2019. [Online]. Available: https://www.ibm.com/cloud/learn/faas.

[2] Rania Mohamed: Software Development, Microservices & Container Management – Part IV – About making Choices – CaaS Platform 4 as SUSE's empowering of Kubernetes. January 2020. [Online]. Available: https://www.suse.com/c/about-making-choices-caasp-4-as-suses-empowering-of-kubernetes/.

[3] International Business Machines Corporation: What is Containers as a service (CaaS)?. [Online]. Available: https://www.ibm.com/services/cloud/containers-as-a-service.

[4] Rick Villars, Chris Barnard, Vice President, Jennifer Cooke, Glen Duncan, Susan G. Middleton, Archana Venkatraman: Worldwide Datacenter 2020 Predictions. October 2019. [Online]. Available: https://www.idc.com/research/viewtoc.jsp?containerId=US44747919.

[5] Shripad Nadgowda, Sahil Suneja, Nilton Bila, Canturk Isci: Voyager: Complete Container State Migration. ICDCS 2017: 2137-2142.

[6] Yuan-Yao Shih, Hsin-Peng Lin, Ai-Chun Pang, Ching-Chih Chuang, Chun-Ting Chou: An NFV-Based Service Framework for IoT Applications in Edge Computing Environments. IEEE Trans. Netw. Serv. Manag. 16(4): 1419-1434 (2019).

[7] Chirag Barhate: Edge Computing is the Future, and Containerization is that Indispensable First Step. May 2020. [Online]. Available: https://cloudhedge.io/edge-computing-is-the-future-and-containerization-is-that-indispensable-first-step/.

[8] The Growth in Connected IoT Devices Is Expected to Generate 79.4ZB of Data in 2025, According to a New IDC Forecast. June 2019. [Online]. Available: https://www.businesswire.com/news/home/20190618005012/en/The-Growth-in-Connected-IoT-Devices-is-Expected-to-Generate-79.4ZB-of-Data-in-2025-According-to-a-New-IDC-Forecast.

[9] Google: Containers at Google. [Online]. Available: https://search.iwiki.uk/extdomains/cloud.google.com/containers.

[10] Domenico Cotroneo, Roberto Natella, Roberto Pietrantuono, Stefano Russo: Software Aging Analysis of the Linux Operating System. ISSRE 2010: 71-80.

[11] Hiroyuki Okamura, Chao Luo, Tadashi Dohi: Estimating Response Time Distribution of Server Application in Software Aging Phenomenon. ISSRE (Supplemental Proceedings) 2013: 281-284.

[12] Chafika Benzaid, Tarik Taleb: AI-Driven Zero Touch Network and Service Management in 5G and Beyond: Challenges and Research Directions. IEEE Network 34(2): 186-194 (2020).

[13] Near-zero downtime: Overview and trends. [Online]. Available: https://www.reliableplant.com/Read/6971/downtime-trends.

[14] Hiroyuki Okamura, Jungang Guan, Chao Luo, Tadashi Dohi: Quantifying Resiliency of Virtualized System with Software Rejuvenation. IEICE Trans. Fundam. Electron. Commun. Comput. Sci. 98-A(10): 2051-2059 (2015).

[15] Matheus Torquato, Erico A. C. Guedes, Paulo R. M. Maciel, Marco Vieira: A Hierarchical Model for Virtualized Data Center Availability Evaluation. EDCC 2019: 103-110.

[16] Matheus Torquato, Paulo Maciel, Marco Vieira: Availability and Reliability Modeling of VM Migration as Rejuvenation on a System under Varying Workload. Software Quality Journal: 1-25 (2020).

[17] Besmir Tola, Gianfranco Nencioni, Bjarne E. Helvik: Network-Aware Availability Modeling of an End-to-End NFV-Enabled Service. IEEE Trans. Netw. Serv. Manag. 16(4): 1389-1403 (2019).

[18] Besmir Tola, Yuming Jiang, and Bjarne E. Helvik: Model-Driven Availability Assessment of the NFV-MANO with Software Rejuvenation. IEEE Trans. Netw. Serv. Manag. (Early Access) (2021).

[19] Dong-Yuh Yang, Chia-Huang Wu: Evaluation of the availability and reliability of a standby repairable system incorporating imperfect switchovers and working breakdowns. Reliab. Eng. Syst. Saf. 207: 107366 (2021).

[20] Stefano Sebastio, Rahul Ghosh, Tridib Mukherjee: An Availability Analysis Approach for Deployment Configurations of Containers. IEEE Trans. on Serv. Comput. 14(1): 16-29 (2021).

[21] Mario Di Mauro, Giovanni Galatro, Fabio Postiglione, Marco Tambasco: Performability of Network Service Chains: Stochastic Modeling and Assessment of Softwarized IP Multimedia Subsystem. IEEE Trans. on Dependable and Secur. Comput. (Early Access) (2021).

[22] Fumio Machida, Jianwen Xiang, Kumiko Tadano, Yoshiharu Maeno: Lifetime Extension of Software Execution Subject to Aging. IEEE Trans. on Reliab. 66(1): 123-134 (2017).

[23] Jing Bai, Xiaolin Chang, Fumio Machida, Kishor S. Trivedi, Zhen Han: Analyzing Software Rejuvenation Techniques in a Virtualized System: Service Provider and User Views. IEEE Access 8: 6448-6459 (2020).

[24] Jing Bai, Xiaolin Chang, Gaorong Ning, Zhenjiang Zhang, Kishor S. Trivedi: Service Availability Analysis in a Virtualized System: A Markov Regenerative Model Approach. IEEE Trans. on Cloud Comput. (Early Access) (2020).

[25] Jing Bai, Xiaolin Chang, Kishor S. Trivedi, Zhen Han: Resilience-Driven Quantitative Analysis of Vehicle Platooning Service. IEEE Trans. Veh. Technol. 70(6): 5378-5389 (2021).

[26] Jing Bai, Xiaolin Chang, Fumio Machida, Lili Jiang, Zhen Han, Kishor S. Trivedi: Impact of Service Function Aging on the Dependability for MEC Service Function Chain. IEEE Tans. Dependable Secur. Comput. (Early Access) (2022).

[27] Aiguo Xie, Peter A. Beerel: Efficient State Classification of Finite-state Markov Chains. IEEE Trans. on CAD of Integrated Circuits and Systems 17(12): 1334-1339 (1998).

[28] Felipe Oliveira, Jean Araujo, Rúbens de Souza Matos Júnior, Luan Lins, André Rodrigues, Paulo R. M. Maciel: Experimental Evaluation of Software Aging Effects in a Container-Based Virtualization Platform. SMC 2020: 414-419.

[29] Tetiana Markova: Live Migration of Docker Containers within Cloud Regions. January 2016. [Online]. Available: https://jelastic.com/blog/live-migration-of-docker-containers-within-cloud-regions/.

[30] Gregory Levitin, Liudong Xing, Hanoch Ben-Haim: Optimizing software rejuvenation policy for real time tasks. Reliab. Eng. Syst. Saf. 176: 202-208 (2018).

[31] Gregory Levitin, Liudong Xing, Hong-Zhong Huang: Optimization of partial software rejuvenation policy. Reliab. Eng. Syst. Saf. 188: 289-296 (2019).

[32] Gregory Levitin, Liudong Xing, Yanping Xiang: Cost Minimization of Real-time Mission for Software Systems with Rejuvenation. Reliability Engineering & System Safety 193: 106593 (2020).

[33] Kishor S. Trivedi, Andrea Bobbio. Reliability and Availability Engineering: Modeling, Analysis, and Applications. Cambridge University Press, 2017.

[34] Ryan Florin, Aida Ghazizadeh, Puya Ghazizadeh, Stephan Olariu and Dan C. Marinescu. Enhancing Reliability and Availability through Redundancy in Vehicular Clouds. IEEE Trans. on Cloud Comput. (2019).

[35] GaoRong Ning, Jing Zhao, Yunlong Lou, Javier Alonso, Rivalino Matias, Kishor S. Trivedi, BeiBei Yin, KaiYuan Cai: Optimization of Two-Granularity Software Rejuvenation Policy Based on the Markov Regenerative Process. IEEE Trans. Reliab. 65(4): 1630-1646 (2016).

[36] Besmir Tola, Gianfranco Nencioni, Bjarne E. Helvik, Yuming Jiang: Modeling and Evaluating NFV-Enabled Network Services under Different Availability Modes. DRCN 2019: 1-5.

[37] Besmir Tola, Yuming Jiang, Bjarne E. Helvik: On the Resilience of the NFV-MANO: An Availability Model of a Cloud-native Architecture. DRCN 2020: 1-7.

[38] Andrew Machen, Shiqiang Wang, Kin K. Leung, Bongjun Ko, Theodoros Salonidis: Live Service Migration in Mobile Edge Clouds. IEEE Wirel. Commun. 25(1): 140-147 (2018).

[39] Maplesoft, Inc., Maple 18, http://www.maplesoft.com/products/maple.

[40] Javier Alonso, Jordi Torres, Josep Lluis Berral, Ricard Gavaldà: Adaptive On-Line Software Aging Prediction based on Machine Learning. DSN 2010: 507-516.

[41] Abderrahmane Boudi, Ivan Farris, Miloud Bagaa, Tarik Taleb: Assessing Lightweight Virtualization for Security-as-a-Service at the Network Edge. IEICE Trans. Commun. 102-B(5): 970-977 (2019).

[42] Domenico Cotroneo, Roberto Natella, Roberto Pietrantuono, Stefano Russo: A Survey of Software Aging and Rejuvenation Studies. JETC 10(1): 8:1-8:34 (2014).

[43] Wai-Leong Yeow, Cédric Westphal, Ulas C. Kozat: Designing and Embedding Reliable Virtual Infrastructures. Comput. Commun. Rev. 41(2): 57-64 (2011).